\documentclass[12pt]{article}
\usepackage[utf8]{inputenc}

\usepackage{amsmath}
\usepackage{amsfonts}
\usepackage{amssymb}
\usepackage{makeidx}
\usepackage{graphicx}
\usepackage{subcaption}
\usepackage{listings}

\usepackage{apacite}
\usepackage{natbib}

\usepackage{indentfirst}
\usepackage{caption}

\usepackage{ulem}
\usepackage{epsfig}

\usepackage{siunitx}
\usepackage{booktabs}
\newcolumntype{Y}{>{\centering\arraybackslash}X}

\usepackage{graphicx,psfrag,epsf}
\usepackage{enumerate}

\usepackage{url} 
\usepackage{newtxmath}  
\usepackage{setspace}
\usepackage{bbm}
\usepackage{dsfont}
\usepackage{newtxtext}      
\usepackage{url}

\usepackage{authblk}
\usepackage[pdfborder={0 0 0}]{hyperref}

\title{\bf A Rational Finance Explanation of the Stock Predictability Puzzle}
\author[a]{Abootaleb Shirvani}
\author[b]{Svetlozar T. Rachev}
\author[c]{Frank J. Fabozzi}

\begin{spacing}{.7}
\affil[a]{\small Department of Mathematics and Statistics, Texas Tech University\\
\url{abootaleb.shirvani@ttu.edu}}
\affil[b]{\small  Department of Mathematics and Statistics, Texas Tech University\\
\url{zari.rachev@ttu.edu}}
\affil[c]{\small EDHEC Business School\\
\url{frank.fabozzi@edhec.edu}}
\end{spacing}

\begin{document}
\thispagestyle{plain}

\date{}

\maketitle
\begin{spacing}{1.00}
\noindent \textbf{Abstract}\ \ \ \ \ 	In this paper, we address one of the main puzzles in finance observed in the stock market by proponents of behavioral finance: the stock predictability puzzle. We offer a statistical model within the context of rational finance which can be used without relying on behavioral finance assumptions to model the predictability of stock returns. 
We incorporate the predictability of stock returns into the well-known Black-Scholes  option pricing formula.
Empirically, we analyze the option and spot trader's market predictability of stock prices by defining a forward-looking measure which we call ``implied excess predictability''. The empirical results indicate
the effect of option trader's predictability of stock returns on the price of stock options is an increasing function of moneyness, while this effect is decreasing for spot traders. These empirical results indicate potential asymmetric predictability of stock prices by spot and option traders. 
We show in pricing options with the strike price significantly higher or lower than the stock price, the predictability of the underlying stock's return  should be incorporated into the option pricing formula. 
In pricing options that have moneyness close to one, stock return predictability is not incorporated into the option pricing model because stock return predictability is the same for both types of traders. In other words, spot traders and option traders are equally informed about the future value of the stock market in this case.
Comparing different volatility measures, we find  that the difference between implied and realized variances or variance risk premium can potentially be used as a stock return predictor.
\\
\\
\noindent \textbf{Keywords}\ \ \ \ \   Predictability of stock returns; behavioral finance; rational dynamic stock pricing theory; option pricing; Stratonovich integral.
\end{spacing}

\begin{spacing}{1.00}
\newpage

\section{Introduction}

In an efficient market, price discovery should be instantaneous and contemporaneous.\footnote{See \cite{Kumar:2013}.} Empirical evidence suggests that the excess aggregate stock market returns are predictable. Using monthly, real, equal-weighted New York Stock Exchange returns from 1941–1986, \cite{Fama:1988} found that the dividend–price ratio can explain 27\% of the variation of cumulative stock returns over the subsequent four years. \cite{Campbell1988a}  specify econometric models of dividend discounting that imply that price dividend ratios predict stock returns. These two studies were among the first to identify this as the ``stock predictability puzzle.”

There are a good number of more recent empirical studies that have investigated the predictability of stock returns.  Some believe stock return predictability is attribute to changes in business conditions, while others attribute it to market inefficiency. 

The majority of work on the predictability of stock returns  is based on statistical, macro, and  fundamental factor analyses models.\footnote{See for example, \cite{Kandel:1996}, \cite{Neely:2000}, \cite{Malkiel:2003}, \cite{Barberis:2003}, \cite{Shiller:2003}, \cite{Avramov:2003}, \cite{Wachter:2009}, \cite{Pesaran:2010}, \cite{Zhou:2010}, and \cite{Bekiros:2013}. The models that have been used are (1) Conditional Capital Asset Pricing Model, (2) vector autoregressive models,  (3)  Bayesian  statistical factor analysis,  (4)  posterior moments of the predictable regression coefficients,  (5)  posterior odds,  (6)  the information in stock prices,  (7)  business cycles effects,   (8)  stock  predictability  of future returns from initial dividend yields,  (9)  firm characteristics as stock return predictors,  (10)  anomalies,  (11)  predictive power of scaled-price ratios such as book-to-market and earnings-to-price, forward spread, and short rate,  (12)  variance risk premia and variance spillovers,  (13)  momentum, market memory and reversals, and  (14)  early announcements and others.}
Recently, a good numbers of studies in behavioral finance have examined behavioral factors that could lead to the predictability of stock returns. 
The behavioral factors that proponents of behavioral finance have suggested that can lead to stock return predictability are (1) sentiment, (2) overconfidence, (3) optimism and wishful thinking, (4) conservatism, euphoria and gloom, (5) self-deception, (6) cursedness, (7) belief perseverance, and (8) anchoring.\footnote{See \cite{Lewellen:2000}, \cite{Barberis:2003}, \cite{Ferson:2006}, \citet[Chapter~1]{Peleg:2008}, and \cite{Daniel:2015}.}

Motivated by the empirical findings that stock returns are predictable,  some researchers have investigated the impact of stock return predictability on the prices of related assets. \cite{Lo:1995}, for example, discussed the effect  of stock return predictability on the price of stock options. They showed that even a small amount of predictability could  have a significant impact on option pricing. \cite{Liao:2006} demonstrated that the effect  of autocorrelated returns on European option prices is significant. 
\cite{Huang:2009} and \cite{Paschke:2010}  offer even more recent examples of studies  about the impact of stock returns predictability on the valuation of options. The upshot of these studies is that to obtain more realistic stock prices, it is essential 
to model and analyze  stock return predictability and incorporate its impact into stock log-returns and option pricing models.

Modeling and analyzing the stock return predictability is  crucial for stock  and risk managers.\footnote{See \cite{Shirvani:2019}.} \cite{Lo:1995} introduced a model to price options when stock returns are predictable. Their model is based on a specially designed multivariate trending Ornstein–Uhlenbeck (O-U) process includes many parameters. The trending O-U processes with small dimensions such as  univariate and bivariate processes are not realistic as noted by  \cite{Lo:1995}.  Moreover, in their model, predictability is induced by the drift parameter, which is not a parameter in the classical Black-Scholes model.

In this paper, we propose a method to model the prediction of stock prices by adjusting the stock predictability as a parameter with the \cite{Black:1973} and \cite{Merton1973a} model framework by using the Stratonovich integral.\footnote{See  \citet[Chapter~2]{Kloeden:2000}, \citet[Chapter~5]{Oksendal:2003}, and \cite{Syga:2015}.} In our model, predictability is viewed as the dividend yield, which we refer to \textit{dividend yield due to predictability}, and is incorporated into the option pricing formula.  We derive an option pricing model by  incorporating the predictable stock returns within the classical Black-Scholes-Merton (BSM) framework. 

Next, we define \textit{implied excess predictability} to compare an option trader's predictability of stock returns with that of a  trader in the cash market (i.e., spot trader). Using the observed price of European call options based on the SPDR S\&P 500 ETF (SPY), we plot the implied excess predictability against ``moneyness'' and time to maturity. The pattern of the implied excess predictability surface shows that at each maturity, an option trader's predictability of the SPY is an increasing function of moneyness. The turning point of the surface is where the moneyness is close to $0.95$. The effect of option trader's predictability of stock returns on the price of stock options increases when the moneyness increases from  $0.95$ to $1.20$.  Conversely,,  when the moneyness decreases from  $0.95$ to $0.7$,
the effect of spot trader's predictability of stock returns on the price of stock options decreases. These empirical results indicate potential asymmetric predictability of stock prices by spot and option traders. 

We demonstrate that in pricing an option with significant intrinsic value, stock return predictability should be incorporated into the BSM model. In pricing options that have moneyness close to one stock predictability is not incorporated into the BSM model because stock predictability is the same for both types of traders. In other words, spot traders and option traders are equally informed about stock market in this case. We show a popular stock market volatility index –- the CBOE volatility index (VIX \footnote{VIX is an index created by CBOE, representing 30-day implied volatility calculated by S\&P500 options, see {http://www.cboe.com/vix.}}) --  is potentially more informative than the other volatility measures (historical, realized, and time series estimation method volatility) for predicting stock returns. The variance risk premium -- the difference between implied variance and realized variance -- can potentially  predict stock market returns.

This paper is organized as follows. The next section describes our methodology for modeling the prediction of stock prices. Then we derive an option pricing formula by incorporating the predictability of stock returns into the model. Section 3 describes the results of our model using the S\&P 500 index options. We then analyze and compare the prediction of stock market returns by option and spot traders. Section 5 summarizes our findings.

\section{The Predictability of Stock pricing}
\noindent  A major issue raised by the proponents of behavioral finance is that prices are often predictable.\footnote{See, for example, \cite{Daniel:2015}.}. More precisely, given a stochastic basis $\left(\mathrm{\Omega },\mathcal{F},\mathbb{F}\mathrm{=}\left({\mathcal{F}}_t,t\ge 0\right)\mathrm{,}\mathbb{P}\mathrm{\ }\right)$ a price process $S\left(t\right)$ $t\ge 0$, defined on $\left(\mathrm{\Omega }\mathrm{,}\mathcal{F}\mathrm{,}\mathbb{P}\right)$  is not necessarily $\mathbb{F}$-adapted, it is adapted to an augmented filtration ${\mathbb{F}}^{\left(\mathrm{*}\right)}\supset \bigcup_{t\ge o}{{\mathcal{F}}_t}$, with ${\mathbb{F}}^{\left(\mathrm{*}\right)}\mathrm{\subset }\mathcal{F}$. 

Admitting the fact that stock returns are predictable, we propose a method to model the prediction of stock returns by adjusting the predictability of stock returns.  Our option pricing model is close to the idea put forth by \cite{Shiller:2003} of  ``smart money versus ordinary investors.''
To model the predictability of stock prices, we use the Stratonovich integral\footnote{See \citet[Chapter~2]{Kloeden:2000}, \citet[Chapter~5]{Oksendal:2003}, and \cite{Syga:2015}.}:

\begin{equation}
\label{GrindEQ__1_}
\begin{array} {c}
\int^T_0{\theta \left(t\right){\circ }^{\left(\frac{1}{2}\right)}dB\left(t\right)} 
	\\={\mathrm{lim}}_{0=t^{\left(0\right)}<t^{\left(1\right)}<\dots <t^{\left(k\right)}=T,t^{\left(j\right)}=j\Delta t,\ \Delta t\downarrow 0}\sum^{k-1}_{j=0}{\theta \left(\frac{t^{\left(j+1\right)}+t^{\left(j\right)}}{2}\right)\left(B(t^{\left(j+1\right)})-B(t^{\left(j\right)})\right)}.
\end{array}
\end{equation} 

In \eqref{GrindEQ__1_}, $B\left(t\right)$, $t\ge 0$, is a Brownian motion generating a stochastic basis $\left(\mathrm{\Omega }\mathrm{,}\mathcal{F},\mathbb{F}\mathrm{=}\left({\mathcal{F}}_t,t\ge 0\right)\mathrm{,}\mathbb{P}\mathrm{\ }\right)$, $\theta \left(t\right)$ $t\ge 0$  is $\mathbb{F}$-adapted left-continuous and locally bounded process. An important property of the Stratonovich integral is that it ``looks into the future,'' and therefore, price processes based on the Stratonovich integral possess predictability properties. In sharp contrast, the It\^{o} integral:
\begin{equation}
 \label{GrindEQ__2_} 
	\int^T_0{\theta (t)dB(t)}={\mathrm{lim}}_{0=t^{\left(0\right)}<t^{\left(1\right)}<\dots <t^{\left(k\right)}=T,\,\,t^{\left(j\right)}=j\Delta t,\ \Delta t\downarrow 0}\sum^{k-1}_{j=0}{\theta \left(t^{\left(j\right)}\right)\left(B(t^{\left(j+1\right)})-B(t^{\left(j\right)})\right)\ \ \ \ }\ \ \ \ \  
\end{equation} 
``does not look in the future,'' and thus It\^{o} prices are not predictable.  Combining both integrals \eqref{GrindEQ__1_} and \eqref{GrindEQ__2_} within a Stratonovich $\alpha$-integral with $\alpha \in \left[0,1\right]$ we obtain:
\begin{equation} 
	\label{GrindEQ__3_} 
	\begin{array}{lll}
		\int^T_0{\theta \left(t\right){\circ }^{\left(\alpha \right)}dB\left(t\right)}\\
		={\mathrm{lim}}_{0=t^{\left(0\right)}<t^{\left(1\right)}<\dots <t^{\left(k\right)}=T,t^{\left(j\right)}=j\Delta t,\ \Delta t\downarrow 0}\sum^{k-1}_{j=0}{\theta \left(t^{\left(j\right)}(1-\alpha )+\alpha t^{\left(j+1\right)}\right)\left(B(t^{\left(j+1\right)})-B(t^{\left(j\right)})\right)}\\= 
		2\alpha \int^T_0{\theta \left(t\right){\circ }^{\left(\frac{1}{2}\right)}dB\left(t\right)+}\left(1-2\alpha \right)\int^T_0{\theta \left(t\right)dB\left(t\right)}.
	\end{array}
\end{equation} 

Consider a market with two assets:
$\left(i\right)$ a risky asset (stock)  $\mathcal{S}$ with potentially predictive price process $S\left(t\right)$, $t\ge 0$, following It\^{o} stochastic differential equation (SDE):

\begin{equation}
 \label{GrindEQ__4_} 
 dS\left(t\right)=\mu \left(t,S\left(t\right)\right)dt+\ \sigma \left(t,S\left(t\right)\right)dB\left(t\right),\ t\ge 0,\ S\left(0\right)>0, 
\end{equation} 
where $\mu \left(t,S\left(t\right)\right)=\mu_t \,S\left(t\right)$, and $\sigma \left(t,S\left(t\right)\right)=\sigma_t \,S\left(t\right)$, For the regularity conditions implying existence and uniqueness of the strong solution of \eqref{GrindEQ__3_}, see \citet[Chapter~6]{Duffie:2001}. By the  It\^{o} formula,  stock price dynamics is given by 

\[S\left(t\right)=S\left(0\right){\mathrm{exp} \left\{\int^t_0{\left({\mu }_s-\frac{1}{2}{\sigma }^2_s\right)ds}+\int^t_0{{\sigma }_sdB\left(s\right)}\right\}\ },\ S\left(0\right)>0,\ t\ge 0.\] 

\noindent $\left(ii\right)$ riskless asset (bond) $\mathcal{B}$  with price process $\beta \left(t\right),t\ge 0,\ $ defined by 
\begin{equation} \label{GrindEQ__5_} 
	d\beta \left(t\right)=r_t\beta \left(t\right),r_t=\ r\left(t,S\left(t\right)\right),\ \ \beta \left(0\right)>0, 
\end{equation} 
that is, $\beta \left(t\right)=\beta \left(t\right){\mathrm{exp} \left(\int^t_0{r_sds}\right)\ }t\ge 0.$

Consider a European Contingent Claim (ECC) $\mathfrak{C}$ with price process $\mathcal{C}\left(t\right)=C\left(t,S\left(t\right)\right)$, where $C\left(t,x\right)$, $t\ge 0$, $x>0$, has continuous derivatives  $\frac{\partial C\left(t,x\right)}{\partial t}$ and $\frac{{\partial }^2C\left(t,x\right)}{\partial x^2}$.  $\mathfrak{C}$'s terminal time is $T>0$, and $\mathfrak{C}$ `s terminal payoff is   $\mathcal{C}\left(T\right)=C\left(T,S\left(T\right)\right)=g\left(S\left(T\right)\right),$ for some continuous $g:\left(0,\infty \right)\to R$.

 Assume that a trader ${\beth }^{\left(l\right)}$ takes a long position in $\mathfrak{C}$. Furthermore, when ${\beth }^{\left(l\right)}$ trades stock $\mathcal{S}$ with possibly superior or inferior to \eqref{GrindEQ__4_}, the following Stratonovich $\alpha$ SDE:
\begin{equation}
 \label{GrindEQ__6_} 
	dS\left(t\right)=\mu \left(t,S\left(t\right)\right))dt+\ \sigma \left(t,S\left(t\right)\right){\circ }^{\left(\alpha \right)}dB\left(t\right),\ t\ge 0,\ S\left(0\right)>0, \alpha \in \left[0,1\right].
\end{equation} 

\noindent Thus, the Stratonovich SDE
\[dS\left(t\right)=\mu \left(t,S\left(t\right)\right)dt+\ \sigma \left(t,S\left(t\right)\right){\circ }^{\left(\alpha \right)}dB\left(t\right),\] 
is equivalent to the It\^{o} SDE 
\begin{equation} 
\label{GrindEQ__7_} 
\begin{array}{lllll}
dS\left(t\right)&=&\left(\mu \left(t,S\left(t\right)\right)+\alpha \sigma \left(t,S\left(t\right)\right)\ \frac{\partial \sigma \left(t,S\left(t\right)\right)}{\partial x}\right)dt\ +\sigma \left(t,S\left(t\right)\right)dB\left(t\right)\\
&=&{\mu }^{\left(\alpha \right)}_tS\left(t\right)dt+{\sigma }_tS\left(t\right)dB\left(t\right),\, \,\,\, {\mu }^{\left(\alpha \right)}_t={\mu }_t+\alpha {\sigma }^2_t\,\, ,\, t\ge 0,\ t\ge 0.  
	\end{array}      
\end{equation} 

Assume that a trader ${\beth }^{\left(s\right)}$ takes a short position in $\mathfrak{C}$ trading in the contract where ${\beth }^{\left(l\right)}$ had taken the long position. ${\beth }^{\left(l\right)}$ and ${\beth }^{\left(s\right)}$ have entered the contract $\mathfrak{C}$ as the only participants at the closing bid-ask traded $\mathfrak{C}$-contract.\footnote{We assume that ${\beth }^{\left(l\right)}$ and ${\beth }^{\left(s\right)}$ are the two trading parties in a bid-ask trade of $\mathfrak{C}$ providing the smallest bid-ask spread, which ultimately ends up with the trade transaction of $\mathfrak{C}$.} ${\beth }^{\left(s\right)}$ observes only the dynamics of $\mathcal{S}$ traded by ${\beth }^{\left(l\right)}$ and given by \eqref{GrindEQ__3_}. Furthermore, when ${\beth }^{\left(s\right)}$ trades stock $\mathcal{S},$ with dynamics following Stratonovich $\gamma $ SDE:

\begin{equation}
	\label{GrindEQ__8_} 
	dS\left(t\right)=\mu \left(t,S\left(t\right)\right)dt+\sigma \left(t,S\left(t\right)\right){\circ }^{\left(\gamma \right)}dB\left(t\right),\ t\ge 0,\ S\left(0\right)>0, 
\end{equation} 
for some $\gamma \in \left[0,1\right]$; that is,
\begin{equation} \label{GrindEQ__9_} 
dS\left(t\right)={\mu }^{\left(\gamma \right)}_tS\left(t\right)dt+{\sigma }_tS\left(t\right)dB\left(t\right),\ \ {\mu }^{\left(\alpha \right)}_t={\mu }_t+\gamma {\sigma }^2_t\ t\ge 0,\ t\ge 0.                       
\end{equation} 

The $\mathfrak{C}$-dynamics as traded by ${\beth }^{\left(l\right)}$ is determined by the It\^{o} formula:

\begin{equation}
\label{GrindEQ__10_} 
\begin{array}{lc}
dC\left(t,S\left(t\right)\right)\\=\left\{\frac{\partial C\left(t,S\left(t\right)\right)}{\partial t}+\frac{\partial C\left(t,S\left(t\right)\right)}{\partial x}{\mu }^{\left(\gamma \right)}_tS\left(t\right)+\frac{1}{2}\frac{{\partial }^2C\left(t,S\left(t\right)\right)}{\partial x^2}{\sigma }^2_tS{\left(t\right)}^2\right\}dt+\frac{\partial C\left(t,S\left(t\right)\right)}{\partial x}{\sigma }_tS\left(t\right)dB\left(t\right). 
\end{array}
\end{equation}

To hedge the risky position, ${\beth }^{\left(s\right)}$ forms a replicating self-financing strategy given by the pair  $a\left(t\right),\ b\left(t\right),t\ge 0,$ where  $C\left(t,S\left(t\right)\right)=a\left(t\right)S\left(t\right)+b(t)\beta \left(t\right)$ with $dC\left(t,S\left(t\right)\right)=a\left(t\right)dS\left(t\right)+b\left(t\right)d\beta \left(t\right).$ 
Thus,
\begin{equation}
\label{GrindEQ__11_} 
dC\left(t,S\left(t\right)\right)=\left(a\left(t\right){\mu }^{\left(\gamma \right)}_t+b\left(t\right)r_t\beta \left(t\right)\right)S\left(t\right)dt+a\left(t\right)\sigma \left(t,S\left(t\right)\right)dB\left(t\right) 
\end{equation} 
From \eqref{GrindEQ__10_} and \eqref{GrindEQ__11_}, ${\beth }^{\left(s\right)}$ obtains $a\left(t\right)=\frac{\partial C\left(t,S\left(t\right)\right)}{\partial x}$, and $b\left(t\right)\beta \left(t\right)=C\left(t,S\left(t\right)\right)-\frac{\partial C\left(t,S\left(t\right)\right)}{\partial x}S\left(t\right).$ Equating the terms with $dt$ and setting $S\left(t\right)=x,$ results in  the following partial differential equation (PDE):
\begin{equation}
\label{GrindEQ__12_} 
0=\frac{\partial C\left(t,x\right)}{\partial t}+\frac{\partial C\left(t,x\right)}{\partial x}\left(r_t-p{\sigma }^2_t\right)x-r_tC\left(t,x\right)+\frac{1}{2}\frac{{\partial }^2C\left(t,x\right)}{\partial x^2}{\sigma }^2_t\ x^2,\, \,p=\gamma -\alpha .
\end{equation}
We call $p\in [-1,1]$ the \textit{excess predictability of }$\mathcal{S}$ \textit{traded by} ${\beth }^{\left(s\right)}$ over the $\mathcal{S}$-dynamic, when $\mathcal{S}$ is traded by ${\beth }^{\left(l\right)}$.  In the classical Black-Scholes model, dividends were not accounted for in the model. 
If we assume that the stock $\mathcal{S}$ provides a continuous dividend yield of  $p{\sigma}^2_t$ (i.e., the dividend paid over interval $(t,t+dt]$ equals $p{\sigma}^2_t S_t$)  we obtain the Black-Scholes partial differential equation given by \eqref{GrindEQ__12_}. 
Borrowing this idea, stock with continuously compounded dividend yield $p{\sigma}^2_t$, we denote $D_y\left(t\right)=\ p{\sigma }^2_t$  as the \textit{dividend yield due to predictability}. As the payment of dividends impacts the option price of the underlying stock, the stock return predictability impacts the price of options. Depending on the sign of $p,\ $ $D_y\left(t\right)$ could be positive or negative. When $p=0,$ we obtain the classical Black-Scholes equation.

In particular, $\mathrm{\ }\mathfrak{C}$-price dynamics is given by\footnote{See \citet[Section~6]{Duffie:2001}.}
\begin{equation}
\label{GrindEQ__13_}  \mathcal{C}\left(t\right)={\mathbb{E}}^{\mathbb{Q}}_t\left\{e^{-\int^T_t{r_udu}}g\left(S(T)\right)\}\right\},\ \ \ t\in \left[0,T\right),
\end{equation}
where  $\mathbb{Q}$ is the equivalent martingale measure for the dividend-stock-price. That is, $\mathbb{Q}\mathrm{\sim }\mathbb{P}\mathrm{,}$ and the discounted gain process  $G^{\left(Y\right)}\left(t\right)=X^{\left(Y\right)}\left(t\right)+D^{\left(Y\right)}\left(t\right)$ is a $\mathbb{Q}$-martingale,
\[S\left(t\right)=S\left(0\right){\mathrm{exp} \left\{\int^t_0{\left({\mu }_s-\frac{1}{2}{\sigma }^2_s\right)ds}+\int^t_0{{\sigma }_sdB\left(s\right)}\right\}\ }, S\left(0\right)>0,\ t\ge 0.\]

\noindent $Y\left(t\right)=\frac{1}{\beta \left(t\right)},\ t\ge 0$,  $\ X^{\left(Y\right)}\left(t\right)=X\left(t\right)Y\left(t\right)$, and $dD^{\left(Y\right)}\left(t\right)=Y\left(t\right)dD\left(t\right)$.  The dynamics of $\mathcal{S}$  on $\mathbb{Q}$ is given by $dS\left(t\right)=\left(r_t-D_y\left(t,x\right)\right)S\left(t\right)dt+dB\left(t\right)$, where $r_t$ is the risk-free rate at time $t$.

 In conclusion, with \eqref{GrindEQ__13_} we are able to incorporate the predictability of stock returns into option pricing formula within the classical Black-Scholes-Merton framework.

Suppose  $\mathfrak{C}$  is a European call option with maturity $T$ and strike $K$, and $g\left(S(T)\right)={\mathrm{max} \left(S\left(T\right)-K,0\right)}$. Then for time to maturity, $\tau =T-t$, the value of a call option for a dividend-paying underlying stock in terms of the Black–Scholes parameters is \footnote{See, for example, Hull (2009), Chapter 13.} 
\begin{equation}
\label{GrindEQ__14_}
\mathcal{C}\left(t\right)=c\left(S\left(t\right),\tau ,K,r_t,{\sigma }_t,p\right)=S\left(t\right)e^{-D_y\left(t\right)\tau }\mathrm{\Phi }\left(d_{+\ }\right)-Ke^{{-r}_t\tau }\mathrm{\Phi }\left(d_{-\ }\right),
\end{equation}
where  $D_y\left(t\right)=\ p{\sigma }^2_t$, $\mathrm{\Phi }$ denotes the standard normal cumulative distribution function, and  $
d_{\pm \ }=\frac{{\mathrm{ln} \left(\frac{S\left(t\right)e^{-D_y\left(t\right)\tau }}{Ke^{{-r}_t\tau }}\right)\ }\pm \frac{1}{2}{\sigma }^2_t\tau }{{\sigma }_t\sqrt{\tau }}$.
Given put–call parity, the price of a put option, $\mathcal{P}\left(t\right)$ is
\[\mathcal{P}\left(t\right)=\mathcal{C}\left(t\right)+D_y\left(t\right)-S_t+K e^{-r_t\tau}.\]

\section{Implied dividend yield due to predictability}
In this section, we compare the option and spot trader's predictability of stock returns by defining the \textit{implied excess predictability}. Implied excess predictability is a metric that captures the view of the option and spot trader of the likelihood moves in the stock price. It can be used to predict the of stock price from two perspectives. An important characteristic of implied excess predictability is that it is forward looking. It compares the predictability of markets for the given underlying stock market index from two perspectives. Recall that implied excess predictability is calibrated from the BSM  option price formula.

We denote by $p$ the excess predictability of $\mathcal{S}$ \textit{traded by} ${\beth }^{\left(s\right)}$ over the $\mathcal{S}$-dynamic, when $\mathcal{S}$ is traded by ${\beth }^{\left(l\right)}$. To study ${\beth }^{\left(s\right)}$'s stock return predictability (option trader) compared to ${\beth }^{\left(l\right)}$ (spot trader), we define \textit{implied excess predictability }$p=p\left(\frac{S\left(t\right)}{K},\tau \right)$ as a function of moneyness $\frac{S\left(t\right)}{K}$ and time to maturity $\tau $ as the solution of 
\begin{equation}
\label{GrindEQ__15_}
c\left(S\left(t\right),\tau ,K,r_t,{\sigma }_t,p\right)={\mathcal{C}}^{\left(market\right)}\left(t,\ S\left(t\right),\tau ,K\right),
\end{equation}
where ${\mathcal{C}}^{\left(market\right)}\left(t,\ S\left(t\right),\tau ,K\right)$ 
is the call option prices of SPY\footnote{\url{https://nance.yahoo.com/quote/SPY/options?p=SPY}.}.

We assume that SPY-daily closing prices follow
\begin{equation}
\label{GrindEQ__16_} 
S\left(t\right)=S\left(0\right){\mathrm{exp} \left\{\int^t_0{{\nu }_sds}+\int^t_0{{\sigma }_sdB\left(s\right)}\right\}\ },\ S\left(0\right)>0,\ t=k\Delta t,k\in {\mathbb{N}}_+=\left\{0,1,\dots \right\},
\end{equation}
where ${\nu }_s={\mu }_s-\frac{1}{2}{\sigma }^2_s\ $. Thus, the SPY-daily return series is given by 
\begin{equation}
\label{GrindEQ__17_}
R\left(t+\mathrm{\Delta }t\right)={\mathrm{ln} \left(\frac{S\left(t+\Delta t\right)}{S\left(t\right)}\right)\ }=\int^{t+\Delta t}_t{{\nu }_sds}+\int^{t+\Delta t}_t{{\sigma }_sdB\left(s\right)},\ t=k\Delta t,k\in {\mathbb{N}}_+\, .
\end{equation}
Thus, on $\mathbb{Q}$, the SPY daily return is given by $dS\left(t\right)=\left(r_t-D_y\left(t,x\right)\right)S\left(t\right)dt+dB\left(t\right)$. The value of a call option for the time to maturity, $\tau =T-t$,
is given by \eqref{GrindEQ__14_}. We calculate the implied excess predictability by taking the option's market price, entering it into the \eqref{GrindEQ__15_} formula, and back-solving for the value of $p$.

Here, we compare the option and spot trader's predictability of stock returns by using the implied excess predictability. Rather than looking at individual stocks, our analysis will focus on the aggregate stock market. In our case, the SPY is the proxy we use for the aggregate stock  market. We compare the predictability of markets for the given underlying stock market index from two perspectives, by doing so it provides important insight about the view of option and spot traders regarding the future price of the stock market.

We use call option prices  from $01/02/2015$ to $01/10/2015$ with different expiration dates and strike prices. The expiration date varies from $1/2/2015$
to $6/20/2015$, and the strike price varies from $80$ to $250$ among different call option contracts. The midpoint of the bid and ask is used in the computation. As the underlying of the call option, the SPY index price was $206.38$ on $01/02/2015$. We use the 10-year Treasury yield curve rate \footnote{https://www.treasury.gov/.} on $01/02/2015$ as the risk-free rate $r_t$, here $r_t = 0.0212$. 

As an estimates for ${\sigma }_t$, we use the following four metrics: (1) daily closing values of  $VIX_t/\sqrt{365}$; (2) historical volatility based on one-year historical data; 
(3) realized volatility over one-year historical data; and (4) estimated volatility over one-year by modeling time series with classical methods ARIMA$(1,1)$-GARCH$(1,1)$ with the Student’s $t$ distribution as an innovation distribution.  The minimum estimated value for ${\sigma }_t$ is derived where the realized volatility is applied and the maximum estimated value is derived where the daily closing values of VIX is used .  

\begin{figure}[htb!]
	\centering
	\includegraphics[scale=0.3]{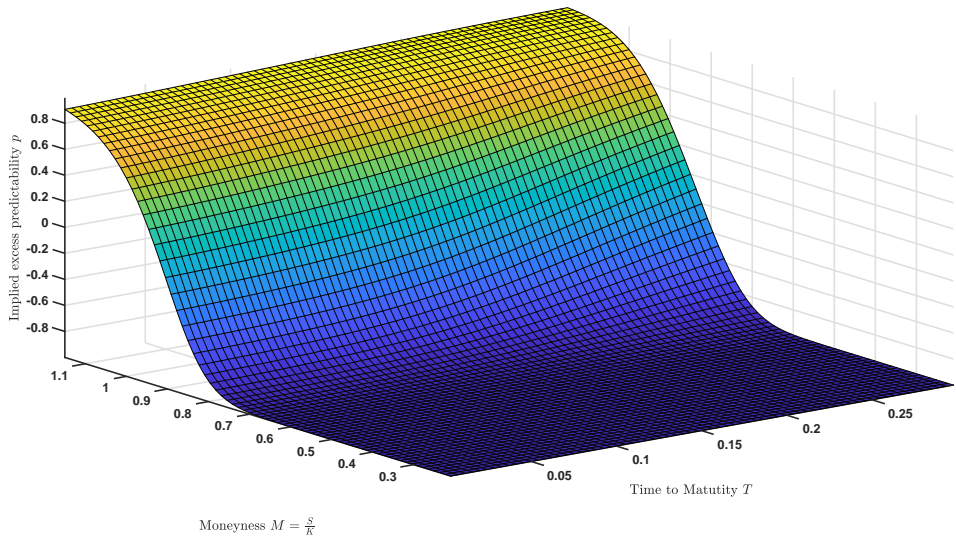}
	\caption{Implied dividend yield against time to maturity and moneyness.}
	\label{Implied_excess_Pre}
\end{figure}

Since implied excess predictability surfaces of all models are very similar, we plot the excess predictability surface when ${\sigma }_t$ is estimated from realized volatility. The implied excess predictability surface is graphed against both a standard measure of ``moneyness'' and time to maturity (in year) in Figure \ref{Implied_excess_Pre}. Recall that a high value for $p$ (close to one) means excess predictability of SPY daily return traded by ${\beth }^{\left(s\right)}$ over the predictability of SPY traded by ${\beth }^{\left(l\right)}$. In other words, $p=1$ means  that option traders potentially predict the future of the SPY returns better than the spot trader. The opposite is true when $p=-1$. Recall that the implied excess predictability surface is an increasing function of ${\sigma }_t$. 

\begin{figure}[htb!]
	\centering
	\includegraphics[scale=0.3]{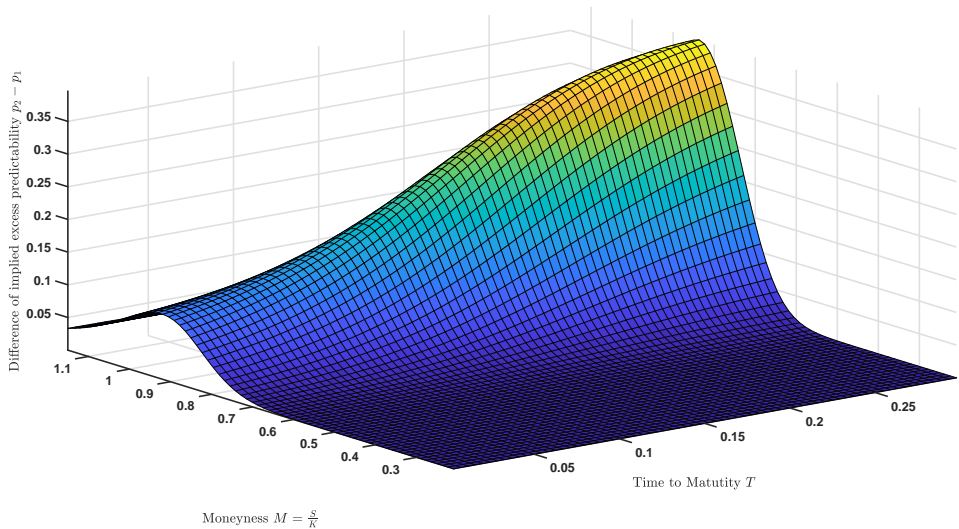}
	\caption{Relative difference of excess predictability VIX-Realized model against time to maturity and Moneyness.}
	\label{VIX_Realized}
\end{figure} 

Figure \ref{Implied_excess_Pre} indicates that at each maturity, implied excess predictability of option traders increase as moneyness increases. Where the moneyness varies in $(0,0.7)$, the surface is flat at point $-1$, indicating higher predictability of spot traders comparing to option traders. Thus, to price significant out-the-money options, the value of $p$ in the model should be $-1$.  
Where the moneyness varies in $(1.05,1.15)$, in-the-money options, the value of $p$ starts increasing from $0.5$, and flats out at point $1$. This finding indicates that option traders can potentially predict market changes better than spot traders when the option is in-the-money. In this case, for pricing in-the-money option, the value of $p$ in the log-return model should be $1$.  

The turning point of the surface is where the moneyness is close to $0.95$. When the moneyness varies in $(0.90,1.05)$,  $p$ varies in $(-0.5,0.5)$. This is the range that spot and option traders are equally informed about the market, and the predictability of the market is equal for both traders. Thus, to price options with no significant intrinsic value, the classical BSM equation can be used.

\begin{figure}[htb!]
	\centering
	\includegraphics[scale=0.3]{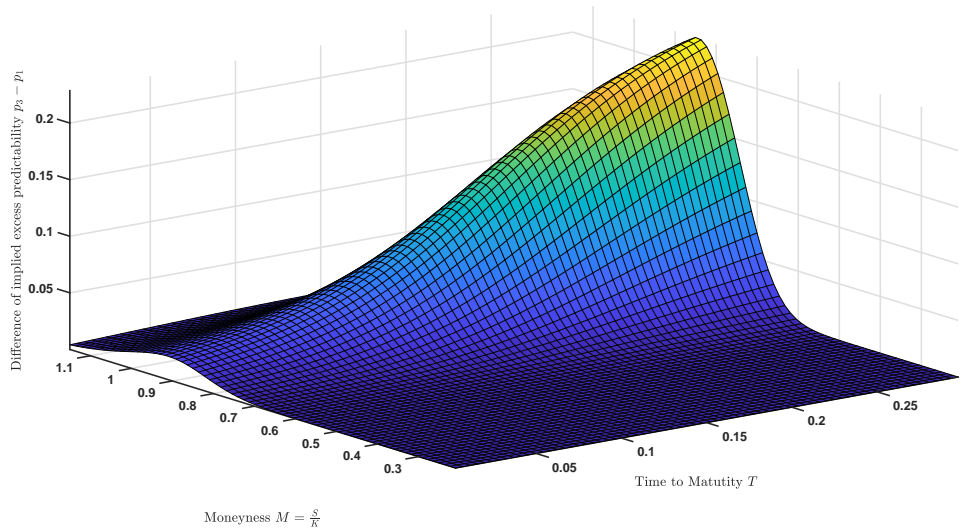}
	\caption{Relative difference of excess predictability Historical-Realized model against time to maturity and Moneyness.}
	\label{Historical_Realized}
\end{figure}

\begin{figure}[htb!]
	\centering
	\includegraphics[scale=0.3]{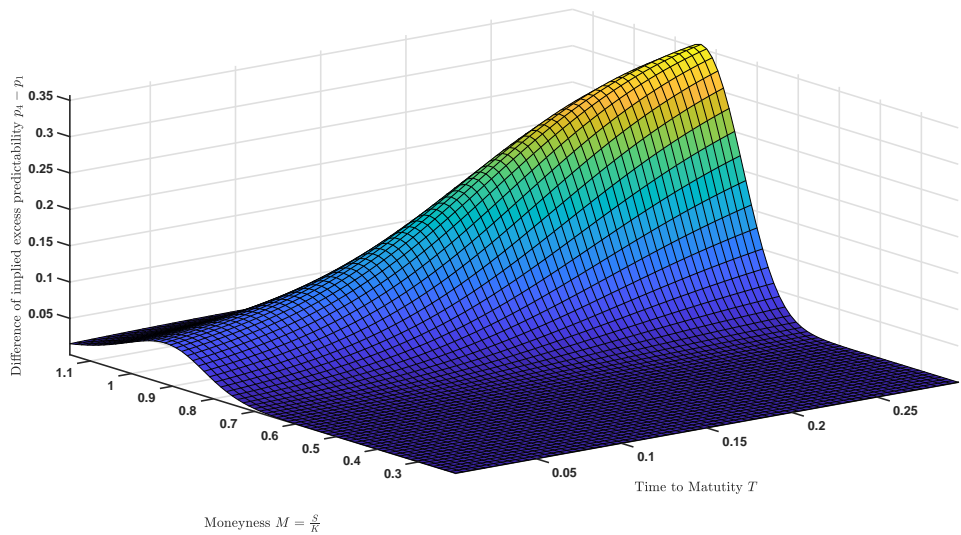}
	\caption{Relative difference of excess predictability time series-Realized model against time to maturity and Moneyness.}
	\label{Garch_Realized}
\end{figure}

As we mentioned, the other four surfaces are very similar. Here, instead of plotting the four similar surfaces, we plot the relative difference of the excess predictability of each surface to the surface derived from realized variance, denoted by $p_i-p_1$, where $i=2,3,4$. Here (1) $p_2$ refers to the excess predictability surface when $\sigma_t$ is imputed from the VIX index, (2) $p_3$ is where $\sigma_t$ imputed by historical volatility, and (3) $p_4$ is where $\sigma_t$ is estimated by time series models.
Figures \ref{VIX_Realized}-\ref{Garch_Realized} show the relative difference of excess predictability for each surface. In all surfaces, where the moneyness varies in $(0.90,1.05)$, the relative difference is significant. At each value for moneyness in $(0.90,1.05)$, the relative difference of excess predictability increases as time to maturity increases.

\cite{Zhou:2010} defined variance risk premium  at time $t$ as the difference between the ex-ante risk-neutral expectation and the objective or statistical expectation of the return variance over the $[t, t + 1]$ time interval,
\begin{eqnarray}
\label{VRP}
VAR_t=E_{t}^{Q}(Var(r_{t+1}))-E_{t}^{P}(Var(r_{t+1})),
\end{eqnarray} 
which is not directly observable in practice. In practice, the
risk-neutral expectation of the return variance, $E_{t}^{Q}(Var(r_{t+1}))$, is typically replaced by the VIX$^2$ index and  statistical expectation of the return variance, $E_{t}^{P}(Var(r_{t+1}))$,  is estimated by realized variance. 

\cite{Zhou:2010} showed that the difference between implied variance and realized variance ((i.e., variance risk premium) can be used for the short-term predictability of equity returns, bond returns, forward premiums, and credit spreads. 
Comparing Figures \ref{VIX_Realized}-\ref{Garch_Realized}, the most significant relative difference of excess predictability is observed in Figure \ref{VIX_Realized}. It indicates that the VIX index contains more information about the stock market compared to the other metrics. Figure \ref{Historical_Realized}, the historical-realized surface, has the minimum relative difference. 

Recall that the maximum and minimum values for $\sigma_t$ are derived from the VIX and realized volatility. As we observed, the most significant relative difference of excess predictability, Figure \ref{VIX_Realized}, is derived where the VIX and realized volatility are used in the model. Thus, by comparing equation \eqref{VRP} for different volatility measures and the fact that excess predictability is an increasing function of $\sigma_t$, suggests that the variance risk premium measure potentially contains more information compared to the other variance measures for predicting stock market returns. The historical volatility measure is the poorest metric.


\section{Conclusion}

\noindent In this paper, we studied the predictability of stock returns within the framework of rational finance rather than relying on behavioral finance explanations. We proposed a method to model stock returns by incorporating the predictability of stock returns in the model and then deriving an option pricing formula. To compare the predictability of stock returns by option traders and spot traders, we constructed a forward-looking measure that compares the option and spot trader's stock returns predictability which we called the ``implied excess predictability measure.'' The empirical results indicate that to price a significant in-the-money and out-the-money option, the option's and spot trader's predictability of stock returns should be incorporated into the  BSM model.
For options with a small intrinsic values, spot traders and option traders are equally informed, and the predictability of the market is equal for both traders. In this case, the classical BSM model can be used for option pricing without incorporating stock return predictability.  Finally, we showed that the  difference between implied variance and realized variance, which we called variance risk premium, is an informative measure for predicting the market in contrast to other volatility measures.

\normalem

\end{spacing}

\begin{thebibliography}{}
	
	\bibitem[\protect\citeauthoryear{Avramov}{Avramov}{2003}]{Avramov:2003}
	Avramov, D. (2003).
	\newblock Stock return predictability and asset pricing models.
	\newblock {\em Review of Financial Studies\/}~{\em 17}, 699--738.
	
	\bibitem[\protect\citeauthoryear{Barberis and Thaler}{Barberis and
		Thaler}{2003}]{Barberis:2003}
	Barberis, N. and R.~Thaler (2003).
	\newblock A survey of behavioral finance.
	\newblock In G.~Harris and R.~Stulz (Eds.), {\em Handbook of the Economics of
		Finance}, Chapter~18, pp.\  1051--1121. North Holland, Amsterdam: Elsevier
	Science.
	
	\bibitem[\protect\citeauthoryear{Bekiros}{Bekiros}{2013}]{Bekiros:2013}
	Bekiros, S.~D. (2013).
	\newblock Irrational fads, short-term memory emulation, and asset
	predictability.
	\newblock {\em Review of Financial Economics\/}~{\em 22}, 213--219.
	
	\bibitem[\protect\citeauthoryear{Black and Scholes}{Black and
		Scholes}{1973}]{Black:1973}
	Black, F. and M.~Scholes (1973).
	\newblock The pricing of options and corporate liabilities.
	\newblock {\em Journal of Political Economy\/}~{\em 81}, 637--654.
	
	\bibitem[\protect\citeauthoryear{Campbell and Shiller}{Campbell and
		Shiller}{1988}]{Campbell1988a}
	Campbell, J.~Y. and R.~J. Shiller (1988).
	\newblock The dividend-price ratio and expectations of future dividends and
	discount factors.
	\newblock {\em Review of Financial Studies\/}~{\em 1\/}(3), 195--228.
	
	\bibitem[\protect\citeauthoryear{Daniel and Hirshleifer}{Daniel and
		Hirshleifer}{2015}]{Daniel:2015}
	Daniel, K. and D.~Hirshleifer (2015).
	\newblock Overconfident investors, predictable returns, and excessive trading.
	\newblock {\em Journal of Economic Perspectives\/}~{\em 29}, 61--88.
	
	\bibitem[\protect\citeauthoryear{Duffie}{Duffie}{2001}]{Duffie:2001}
	Duffie, D. (2001).
	\newblock {\em Dynamic Asset Pricing Theory, 3rd Edition}.
	\newblock Princeton University Press: Princeton N.J.
	
	\bibitem[\protect\citeauthoryear{Fama and French}{Fama and
		French}{1988}]{Fama:1988}
	Fama, F.~E. and R.~K. French (1988).
	\newblock Dividend yields and expected stock returns.
	\newblock {\em Journal of Financial Economics\/}~{\em 22}, 3--25.
	
	\bibitem[\protect\citeauthoryear{Ferson}{Ferson}{2006}]{Ferson:2006}
	Ferson, W.~E. (2006).
	\newblock Conditional asset pricing.
	\newblock In A.~C. Lee (Ed.), {\em Encyclopedia of Finance}, Chapter~9, pp.\
	376--383.
	
	\bibitem[\protect\citeauthoryear{Huang, Wu, and Wang}{Huang
		et~al.}{2009}]{Huang:2009}
	Huang, Y.~C., C.~W. Wu, and C.~W. Wang (2009).
	\newblock Valuing american options under \uppercase{ARMA} processes.
	\newblock {\em International Research Journal of Finance and Economics\/}~{\em
		28}, 152– 159.
	
	\bibitem[\protect\citeauthoryear{Kandel and Stambaugh}{Kandel and
		Stambaugh}{1996}]{Kandel:1996}
	Kandel, S. and R.~F. Stambaugh (1996).
	\newblock On the predictability of stock returns: An asset-allocation
	perspective.
	\newblock {\em Journal of Finance\/}~{\em 51}, 385--424.
	
	\bibitem[\protect\citeauthoryear{Kloeden, Platen, and Schurz}{Kloeden
		et~al.}{2000}]{Kloeden:2000}
	Kloeden, P.~E., E.~Platen, and H.~Schurz (2000).
	\newblock {\em Numerical Solution of SDE Through Computer Experiments}.
	\newblock Heidelberg: Springer-Verlag.
	
	\bibitem[\protect\citeauthoryear{Kumar and Chaturvedul}{Kumar and
		Chaturvedul}{2013}]{Kumar:2013}
	Kumar, K. and C.~Chaturvedul (2013).
	\newblock Price leadership between spot and futures markets.
	\newblock {\em Journal of Applied Finance and Banking\/}~{\em 3\/}(1), 93--107.
	
	\bibitem[\protect\citeauthoryear{Lewellen}{Lewellen}{2000}]{Lewellen:2000}
	Lewellen, J.~W. (2000).
	\newblock {\em {On the Predictability of Stock Returns: Theory and Evidence}}.
	\newblock Ph.\ D. thesis, William E. Simon Graduate School of Business
	Administration, University of Rochester.
	
	\bibitem[\protect\citeauthoryear{Liao and Chen}{Liao and
		Chen}{2006}]{Liao:2006}
	Liao, S.~L. and C.~C. Chen (2006).
	\newblock The valuation of \uppercase{E}uropean options when asset returns are
	autocorrelated.
	\newblock {\em Journal of Futures Markets\/}~{\em 26\/}(1), 85--102.
	
	\bibitem[\protect\citeauthoryear{Lo and Wang}{Lo and Wang}{1995}]{Lo:1995}
	Lo, A.~W. and J.~Wang (1995).
	\newblock Implementing option pricing models when asset returns are
	predictable.
	\newblock {\em Journal of Finance\/}~{\em 50\/}(1), 87--129.
	
	\bibitem[\protect\citeauthoryear{Malkiel}{Malkiel}{2003}]{Malkiel:2003}
	Malkiel, B.~G. (2003).
	\newblock The efficient market hypothesis and its critics.
	\newblock {\em Journal of Economic Perspectives\/}~{\em 17}, 59--82.
	
	\bibitem[\protect\citeauthoryear{Merton}{Merton}{1973}]{Merton1973a}
	Merton, R.~C. (1973).
	\newblock Theory of rational option pricing.
	\newblock {\em Bell Journal of Economics and Management Science,\/}~{\em 6},
	141--183.
	
	\bibitem[\protect\citeauthoryear{Neely and Weller}{Neely and
		Weller}{2000}]{Neely:2000}
	Neely, C.~J. and P.~Weller (2000).
	\newblock Predictability in international asset returns: A reexamination.
	\newblock {\em Journal of Financial and Quantitative Analysis\/}~{\em 35},
	601--620.
	
	\bibitem[\protect\citeauthoryear{{\O}ksendal}{{\O}ksendal}{2003}]{Oksendal:2003}
	{\O}ksendal, B.~K. (2003).
	\newblock {\em Stochastic Differential Equations: An Introduction with
		Applications}.
	\newblock Heidelberg: Springer-Verlag.
	
	\bibitem[\protect\citeauthoryear{Paschke and Prokopczusk}{Paschke and
		Prokopczusk}{2010}]{Paschke:2010}
	Paschke, R. and M.~Prokopczusk (2010).
	\newblock Commodity derivative valuation with autoregressive and moving average
	components in the price dynamics.
	\newblock {\em Journal of Banking and Finance\/}~{\em 34}, 2742–2752.
	
	\bibitem[\protect\citeauthoryear{Peleg}{Peleg}{2000}]{Peleg:2008}
	Peleg, E. (2000).
	\newblock {\em Three essays on asset pricing, portfolio choice and behavioral
		finance}.
	\newblock Ph.\ D. thesis, Philosophy in Management, University of California.
	
	\bibitem[\protect\citeauthoryear{Pesaran}{Pesaran}{2010}]{Pesaran:2010}
	Pesaran, M.~H. (2010).
	\newblock Predictability of asset returns and the efficient market hypothesis.
	\newblock In A.~Ullah and D.~E. Giles (Eds.), {\em Handbook of Empirical
		Economics and Finance}, Chapter~11, pp.\  281--31. North Holland, Amsterdam:
	Elsevier Science.
	
	\bibitem[\protect\citeauthoryear{Shiller}{Shiller}{2003}]{Shiller:2003}
	Shiller, R.~J. (2003).
	\newblock From efficient market theory to behavioral finance.
	\newblock {\em Journal of Economic Perspectives\/}~{\em 17}, 83--104.
	
	\bibitem[\protect\citeauthoryear{Shirvani, Rachev, and Fabozzi}{Shirvani
		et~al.}{2019}]{Shirvani:2019}
	Shirvani, A., S.~Rachev, and F.~Fabozzi (2019).
	\newblock Multiple subordinated modeling of asset returns.
	\newblock {\em arXiv:1907.12600 [q-fin.MF]\/}~{\em 7}.
	
	\bibitem[\protect\citeauthoryear{Syga}{Syga}{2015}]{Syga:2015}
	Syga, J. (2015).
	\newblock Semimartingale measure in the investigation of
	\uppercase{S}tratonovich-type stochastic integrals and inclusions.
	\newblock {\em Discussiones Mathematicae, Probability and Statistics\/}~{\em
		35}, 7--27.
	
	\bibitem[\protect\citeauthoryear{Wachter and Warusawitharana}{Wachter and
		Warusawitharana}{2009}]{Wachter:2009}
	Wachter, J.~A. and M.~Warusawitharana (2009).
	\newblock Predictable returns and asset allocation: Should a skeptical investor
	time the market?
	\newblock {\em Journal of Econometrics\/}~{\em 148}, 162--178.
	
	\bibitem[\protect\citeauthoryear{Zhou}{Zhou}{2010}]{Zhou:2010}
	Zhou, H. (2010).
	\newblock Variance risk premia, asset predictability puzzles, and macroeconomic
	uncertainty.
	\newblock {\em Annual Review of Financial Economics\/}~{\em 10\/}(1), 481--497.
	
\end{thebibliography}
\end{document}